\documentclass[twocolumn,showpacs,showkeys,preprintnumbers]{revtex4}%
\usepackage{amssymb}
\usepackage{makeidx}
\usepackage{amsmath}
\usepackage{graphicx}
\usepackage{dcolumn}
\usepackage{bm}
\usepackage{amsfonts}
\providecommand{\U}[1]{\protect\rule{.1in}{.1in}}
\begin{document}
\title{Hydrodynamics of electron-hole fluid photogenerated in a mesoscopic
two-dimensional channel}
\author{M. A. T. Patricio$^{1}$, G. M. Jacobsen$^{2}$, M. D. Teodoro$^{2}$, G. M.
Gusev$^{3}$, A. K. Bakarov$^{4}$, and Yu. A. Pusep$^{1}$}

\begin{abstract}
The dynamics of the diffusion flow of holes photoinjected into a mesoscopic
GaAs channel of variable width, where they, together with background
electrons, form a hydrodynamic electron-hole fluid, is studied using
time-resolved microphotoluminescence. It is found that the rate of
recombination of photoinjected holes, which is proportional to the rate of
their flow, decreases when holes pass through the expanded sections of the
channel. In fact, this is the Venturi effect, which consists in a decrease in
the velocity of the fluid in the expanded sections of the pipe. Moreover, a
non-uniform diffusion velocity profile is observed, similar to the parabolic
Hagen-Poiseuille velocity profile, which indicates a viscous hydrodynamic
flow. It is shown that in argeement with a theory, the magnetic field strongly
suppresses the viscosity of the electron-hole fluid. Additional evidence of
the viscous nature of the studied electron-hole fluid is the observed increase
in the recombination rate with increasing temperature, which is similar to the
decrease in the electrical resistance of viscous electrons with temperature.

\end{abstract}
\date{14/02/2024}
\affiliation{$^{1}$S\~{a}o Carlos Institute of Physics, University of S\~{a}o Paulo, PO Box
369,13560-970 S\~{a}o Carlos, SP, Brazil}
\affiliation{$^{2}$Departamento de F\'{\i}sica, Universidade Federal de S\~{a}o Carlos,
13565-905, S\~{a}o Carlos, S\~{a}o Paulo, Brazil}
\affiliation{$^{3}$Institute of Physics, University of S\~{a}o Paulo, 135960-170 S\~{a}o
Paulo, SP, Brazil}
\affiliation{$^{4}$Institute of Semiconductor Physics, 630090 Novosibirsk, Russia}
\startpage{1}
\endpage{102}

\email{pusep@ifsc.usp.br}
\maketitle

\section{Introduction}

For almost three decades, the hydrodynamics of electron systems has continued
to be one of the most attractive problems in solid state physics. The history
of the hydrodynamic approach in condensed matter physics is well described in
\cite{geim2020}. The hydrodynamic theory is based only on the conservation of
mass, momentum, and energy, which implies an uncomplicated interpretation of
the observed response of electrons to an external disturbance. The laws of
hydrodynamics become applicable to clean electron systems, where
electron-electron collisions with conservation of momentum predominate over
scattering of electrons without conservation of momentum (see for review
\cite{fritz2023}). More explicitly, electron systems reveal collective fluid
behavior when the effective Knudsen parameter $\zeta=\tau_{ee}/\tau_{p}\ll1$,
where $\tau_{ee}$ and $\tau_{p}$ are the momentum-conserving electron-electron
collision time and momentum-relaxing scattering time, respectively. One of the
fundamental properties of hydrodynamic systems is the space dependent drift
velocity which leads to a non-uniform hydrodynamic flow. The non-uniformity of
a viscous hydrodynamic systems is manifested itself in a Hagen-Poiseuille flow
with a parabolic diffusion velocity profile \cite{white}. Another
manifestation of the hydrodynamic nature is the Venturi effect, which is a
direct consequence of the mass continuity principle, and it is one of the key
concepts in the continuum hydrodynamics of an inviscid fluid. This effect is
that the pressure of the fluid decreases and therefore the velocity increases
as the fluid flows through the constricted section of the pipe. The Venturi
effect has found a vast number of applications. The issue related to the
electronic Venturi effect was first discussed in \cite{govorov2004}, where the
mechanism of quantum pumping in a Fermi system, similar to classical Bernoulli
pumping, was considered. The experimental evidences of electronic quantum
pumping were obtained in an electron-avalanche amplifier based on GaAs/AlGaAs
heterostructure in \cite{taubert2010,taubert2011} and in graphene field effect
transistor in \cite{huang2023}.

Due to the interplay between momentum-conserving electron-electron and
momentum-relaxing electron-phonon scattering, the hydrodynamic electron flow
condition is satisfied in a narrow temperature window, typically 4--35 K,
where electron viscosity is temperature dependent. In this temperature range,
electrons exhibit strong and weak hydrodynamic regimes, in which internal
friction and hence shear stress are strong or weak, respectively. In the weak
hydrodynamic regime electrons are practically inviscid. The Hagen-Poiseuille
diffusion velocity profile is a fundamental characteristic of viscous flow.

To observe the inhomogeneous nature of the electron flow, it is necessary to
carry out measurements with spatial resolution. To date, measurements with
spatial resolution, which unambiguously testify to the hydrodynamic nature of
the electron flow, have been carried out in graphene using a scanning
carbon~nanotube single-electron transistor \cite{sulpizio2019}, a nanoscale
quantum spin magnetometers \cite{ku2020} and a scanning tunneling
potentiometry \cite{krebs2023} and in semimetal WTe$_{2}$ using a nanoscale
scanning superconducting quantum interference device \cite{Aharon2022}, and in
GaAs epilayer using a scanning tunneling microscopy \cite{braem2018}.

We recently performed scanning microscopy photocurrent measurements to study
the diffusion of holes photoinjected into a high-mobility GaAs quantum well
(QW) where the electrons reveal hydrodynamic properties
\cite{pusep2022,pusep2023}. It was found that the observed diffusion consists
of the diffusion of heavy and light holes occurring in a viscous electron-hole
fluid. In this paper we address our investigation to the dynamics of diffusion
of photogenerated holes in a mesoscopic GaAs channel of variable width. Using
time-resolved photoluminescence scanning microscopy (TRPL), we demonstrate
both of the aforementioned hydrodynamic features: the Venturi effect and the
parabolic like velocity profile. Such measurements make it possible to
directly investigate the non-uniformity of an electron system, which is one of
the most obvious proofs of its hydrodynamic nature. In addition, an unusual
increase in the recombination rate with temperature was found, which is
analogous to a decrease in the electrical resistance of viscous electrons with temperature.

It is worth noting that according to Ref. \cite{dassarma2022}, in GaAs QWs the
effective Knudsen parameter is generically much smaller than unity and
"electrons in high-mobility 2D GaAs are by far the best system for the direct
observation of collective hydrodynamic effects conditions for studying
electron hydrodynamics".

\section{Experimental details}

Here we studied a single GaAs QW 46 nm thick, grown on a (100)-oriented GaAs
substrate by a molecular beam epitaxy. QW barriers were grown in the form of
short-period GaAs/AlAs superlattices. The sheet electron density and the
mobility were measured at the temperature of 1.4 K were 6.7$\cdot$10$^{11}$
cm$^{-2}$ and 2.0$\cdot$10$^{6}$ cm$^{2}$/V$\cdot$s, respectively using a
standard Hall bar structure in Ref.\cite{gusev2021}, where the viscous
character of electron transport was demonstrated in \cite{gusev2021}. Electron
hydrodynamics was observed at temperatures above 10 K, at least up to 100 K
\cite{gusev2023}, while the electron-hole fluid exhibits hydrodynamic behavior
in the temperature window 4 - 30 K \cite{pusep2022}. The energy structure of
the studied here samples is similar to the structure calculated in
\cite{pusep2023}. The electric field built in the barriers spatially separates
the electrons and holes photogenerated in the barriers. As a result,
photogenerated holes tunnel into the quantum well. Holes injected in this way
into GaAs QW lead to the emergence of a multicomponent hydrodynamic fluid
formed by background electrons and photogenerated holes. The studied here
structure consists of the sections width 4, 10 and 50 $\mu$m. The time of
recombination of photogenerated holes with background electrons was measured
using TRPL in various sections at the emission energy of GaAs QWs. In this
work, the same GaAs QW is used to fabricate a mesoscopic channel with section
widths of 4, 10 and 50 $\mu$m.

Scanning TRPL microscopy experiments were performed at the temperatures 4 K
and 25 K using a helium closed cycle cryostat equipped with a superconducting
magnet (Attocube/Attodry1000). These temperatures were chosen in order to
distinguish the effect of electron viscosity. According to \cite{pusep2022},
the maximum electron viscosity is expected at 25 K. TRPL measurements with a
temporal resolution of 100 ps were made using the Pico Quant/LDH Series diode
lasers emitting 80 MHz pulses at 730 nm (1.7 eV) and 440 nm (2.33 eV) with a
pulse duration of 70 ps. Excitation at 1.7 eV generates electron-hole pairs in
the GaAs QW, but not in the barrier, whose gap is about 1.8 eV, while
excitation energy of 2.33 eV leads to electron-hole excitation both in the QW
and in the barrier. The electron-hole pairs photogenerated in the QW rapidly
recombine and do not contribute to the diffusion. At the same time, holes
injected from the barriers form a diffusion flow in the QW channel.
\cite{pusep2023}. Photoluminescence (PL) emission was dispersed by a 75 cm
Andor/Shamrock spectrometer and the PL decay transients were detected by a
PicoQuant Hybrid PMT detector triggered with a time correlated single photon
PicoQuant/PicoHarp 300 counting system. The laser Instrument Response Function
(IRF), which is responsible for the temporal resolution of the setup was
measured as a transient process of reflected laser light measured at the laser
energy and it is shown in Fig.1 (d). Electron-hole pairs are generated by a
laser spot about 1 $\mu$m in size. The spatial resolution of the setup is
determined by the size of the light collection area, which is estimated at
about 10 $\mu$m in the spectral range of GaAs QW radiation due to chromatic aberration.

The scheme of excitation/collection process is shown in Fig. 1(a). Holes
injected into the quantum well at the focus of the lens diffuse to the
boundaries of the light collection area at a velocity determined by the
properties of the electron-hole fluid. The scheme of PL transitions at laser
excitation energies of 1.7 and 2.33 eV and the corresponding PL spectra
measured at a pump power of 7.5 $\mu$W, which excludes sample heating, are
shown in Fig. 1(b) and Fig. 1(c), respectively. They demonstrate the effect of
hole accumulation in the QW upon excitation at a wavelength of 440 nm. In the
n-type QW under study, the shape of the PL spectrum is affected by the
occupation of the states of the valence band by minority photogenerated holes.
Not all electrons below the Fermi energy contribute equally to recombination
with photogenerated holes. Only a small fraction of electrons near the bottom
of the conduction band is available for this process, since at low
temperatures and low pump powers only a small number of hole states at the top
of the valence band are occupied. The shape of the PL spectra measured upon
excitation at a wavelength of 730 nm, which generates electron--hole pairs in
the GaAs QW, shows that the majority of holes are located at the valence band
extremum, which leads to a strong energy maximum near the band edge. At the
same time, excitation at a wavelength of 440 nm in the AlGaAs barriers causes
the accumulation of holes in the QW, which leads to an increase in the number
of high-energy electrons participating in recombination. In this case, the PL
maximum shifts towards higher energies, close to the Fermi energy. All
experimental data presented below were obtained at a laser pump power 7.5
$\mu$W.

\begin{figure}
\includegraphics[width=9cm]{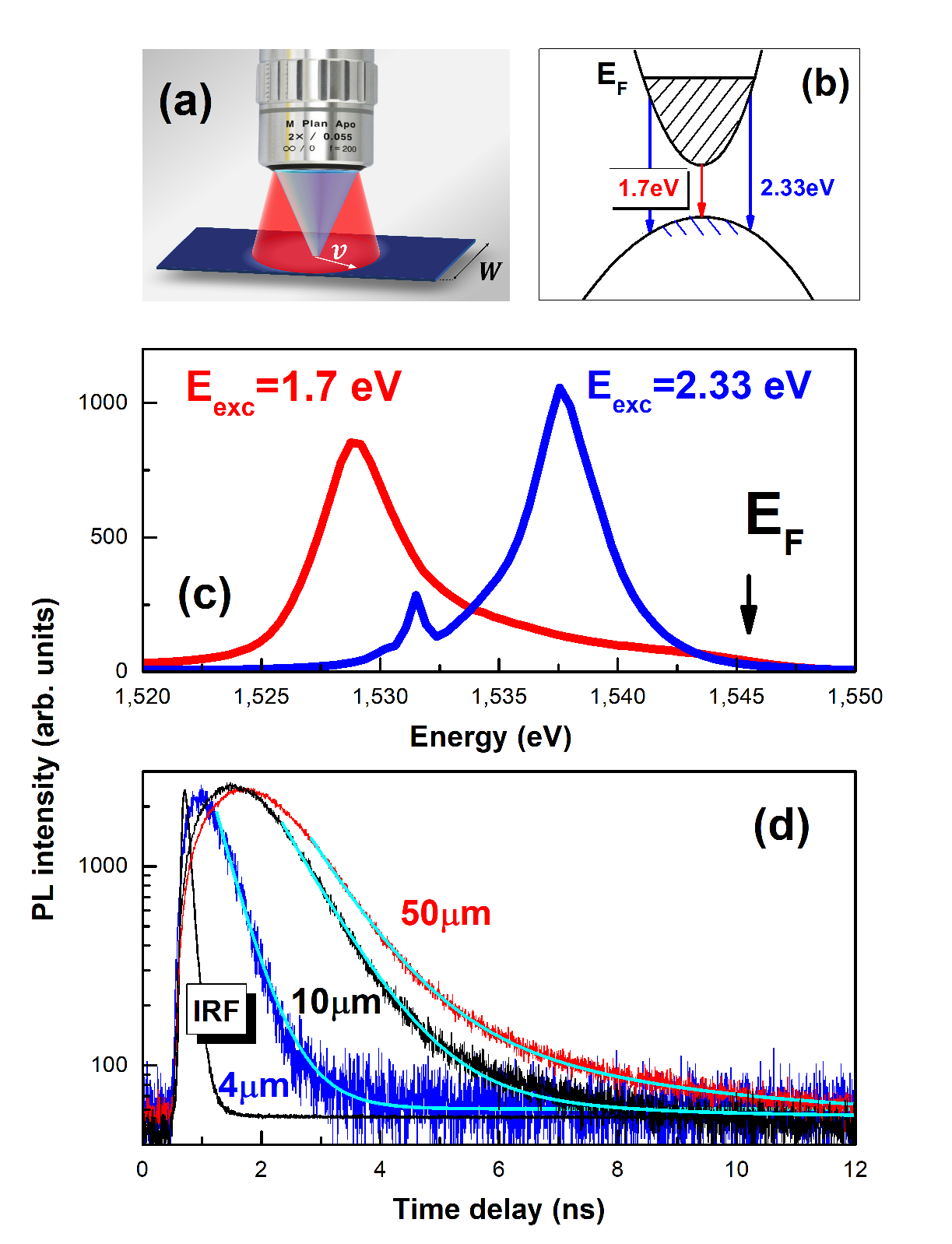}
\caption{(Color online) Sketch of the excitation/collection probe (a), scheme of PL transitions at
different laser excitation energies (b), typical PL spectra
measured at T = 4 K and laser excitation at 440 nm (2.33
eV) and 730 nm (1.7 eV) at the pump power 7.5 W (c),
and typical PL decay transients measured in channel sections of different widths 4, 10 and 50 $\mu$m at T = 4 K and
laser excitation at 440 nm (2.33 eV); the cyan lines show
the results of the best fits (d).}
\end{figure}

Fig. 1 (color on-line). Sketch of the excitation/collection probe (a), scheme
of PL transitions at different laser excitation energies (b), typical PL
spectra measured at T = 4 K and laser excitation at 440 nm (2.33 eV) and 730
nm (1.7 eV) at the pump power 7.5 $\mu$W (c), and typical PL\ decay transients
measured in channel sections of different widths 4, 10 and 50 $\mu$m at T = 4
K and laser excitation at 440 nm (2.33 eV); the cyan lines show the results of
the best fits (d).\bigskip

The sample image is shown in Fig. 2(a). Holes injected from the laser
excitation spot cause a perturbation in the electron hydrodynamic system on
the scale of the hole diffusion length. Due to the hydrodynamic properties of
the electrons, this perturbation propagates over the entire channel. The
diffusion of holes in a system of hydrodynamic electrons can be associated
with the diffusion of particles in an incompressible fluid contained in a
channel. The stationary flow of such particles injected at any point in the
channel creates pressure conditions that depend on the geometry of the
channel, which in turn affect the diffusion rate at the point of injection.

Side contacts connect the channel with contact pads, the area of which
significantly exceeds the area of the channel. Thus, the contact pads serve as
open reservoirs into which photogenerated holes can flow. During measurements,
all contacts remained floating and not grounded, while electrons ensure an
equipotential conditions throughout the entire structure of the sample.

\begin{figure}
\includegraphics[width=9cm]{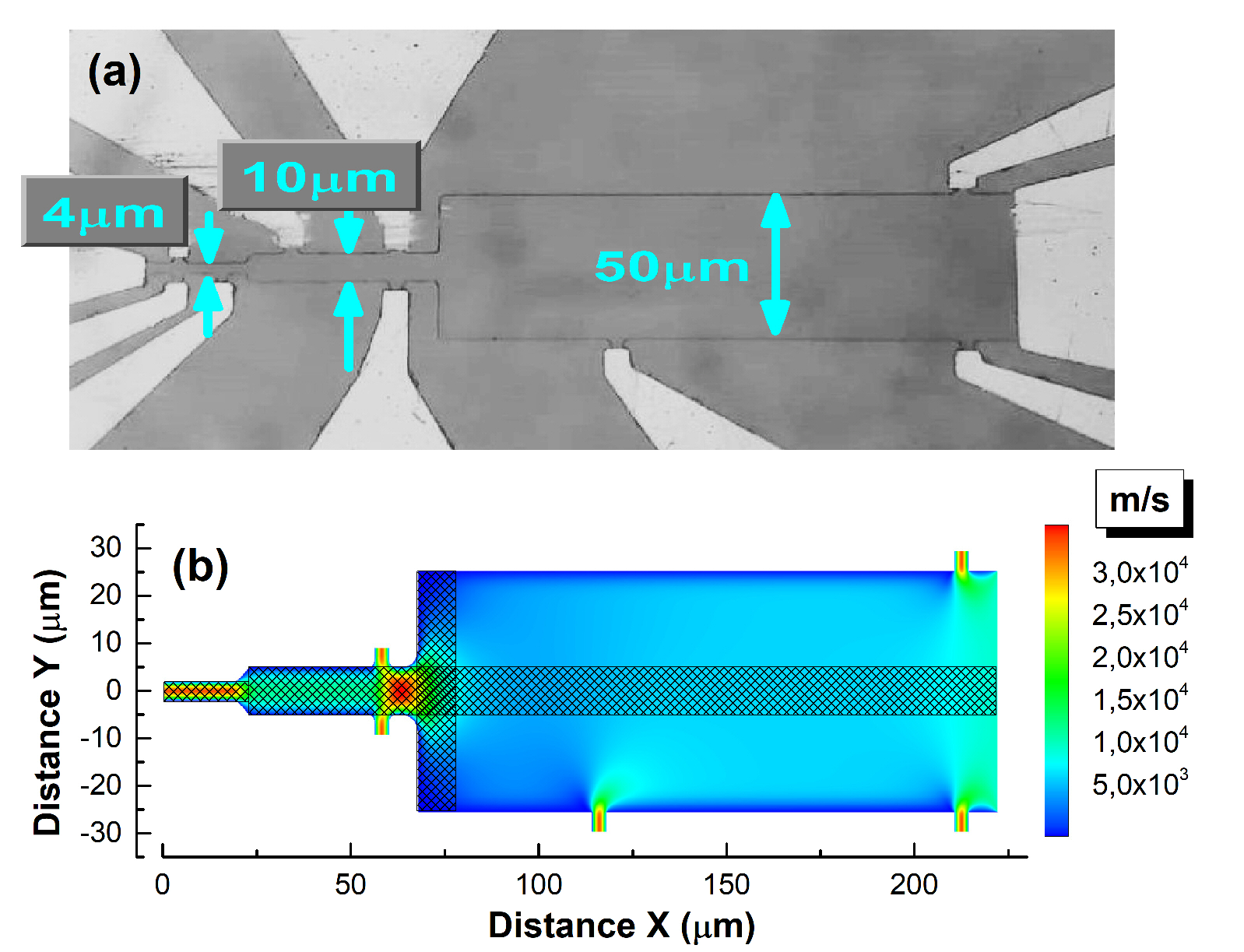}
\caption{(Color online) he sample image (a) and the
velocity flow contour plot in a variable width hydrodynamic channel calculated with the fluid parameters defined in the text (b). The scanning bands parallel and
perpendicular to the channel, determined by the spatial
resolution of the PL, are shown as mesh strips.}
\end{figure}

Fig. 2 (color on-line). The sample image (a) and the velocity flow contour
plot in a variable width hydrodynamic channel calculated with the fluid
parameters defined in the text (b). The scanning bands parallel and
perpendicular to the channel, determined by the spatial resolution of the PL,
are shown as mesh strips.\bigskip\

To model the flow of photogenerated holes in this case, the flow of a
two-dimensional fluid was calculated by the Navier-Stokes equation using the
ANSYS R19.2 code. The parameters required for these calculations were obtained
in similar structures in \cite{pusep2022,pusep2023}. These include: the
diffusion coefficient/kinematic viscosity $D_{hh}=\nu_{hh}=0.17$ m$^{2}$/s and
the diffusion length attributed to heavy holes, measured at T = 25 K, equal to
$L_{hh}$ = $6$ $\mu$m. According to these data the diffusion \ flow velocity
is $v=3D_{hh}/L_{hh}\simeq8.5\cdot10^{4}$ m/s. When simulating hydrodynamic
flow, the contact pins were left open. The corresponding velocity flow contour
plot calculated in viscous fluid is shown in Fig. 2(b). As follows from the
calculations presented in Fig. 2b, the diffusion rate increased as the holes
entered the narrow contact pins, as would be expected from fluid dynamics.

It is important to note that, taking into account the experimental conditions
used, such model calculations provide a qualitative analysis. Nevertheless,
even such a simple model is capable of reproducing the essential features of
the experiment.

To measure the diffusion velocity of holes, we use the fact that the rate of
recombination of photogenerated holes $1/\tau$, where $\tau$ is their
recombination time with background electrons, is proportional to the flow rate
of holes through a certain cross section. In this case the recombination time
is given by $\tau$ = C/$v$, where it is assumed that C = constant at a
definite laser pump power. The recombination times were obtained by fitting
the PL decay transients measured along the scan bands shown in Fig. 2(b).

\section{Results and discussion}

Typical transient of PL decay measured at a laser excitation energy of 2.33
eV, at the maxima of the PL spectra along the center line of channel sections
of different widths are shown in Fig. 1(c). The measured PL transients
exhibited a mostly monoexponential decay with a characteristic time related to
the time of recombination of heavy holes with background electrons. In
contrast to the narrow 14 nm QW studied in \cite{pusep2022,pusep2023}, no
presence of light holes was found in such a wide 46 nm QW. In a narrow QW,
both heavy and light holes participate in recombination process, since spatial
quantization removes the degeneracy of heavy and light holes at the center of
the Brillouin zone. At the same time, in a wide QW, the subbands of heavy and
light holes are close to degeneracy, which leads to the predominant
recombination of heavy holes due to their higher density of states. The data
presented in Fig. 1(d) show that the recombination time decreases with
decreasing channel width, indicating a corresponding increase in the diffusion
velocity expected from the Venturi effect.

To find hydrodynamic evidences in the diffusion flow of photogenerated holes,
the time of their recombination across a wide section of the channel was
measured. The corresponding scanning band, along which the maximum change in
flow velocity is expected, is shown in Fig. 2(b). The recombination times of
photogenerated heavy holes obtained by monoexponential fitting of the
PL\ transients are depicted in Fig. 3(a,b) and demonstrate the non-uniform
velocity distribution over the channel width, which is expected for a
hydrodynamic flow. A significantly stronger change in the recombination time
across the channel width is observed at T = 25 K compared to T = 4 K. This is
due to an increase in the viscosity of the electron-hole fluid at T = 25 K
\cite{pusep2022}. An increase in viscosity leads to an increase in internal
friction and, as a consequence, to increased energy dissipation. The lower the
flow energy, the less its velocity changes when the channel width changes. The
velocity profiles calculated for an inviscid and viscous fluid are shown as
black dotted lines in Fig. 3(a,b). They demonstrate inhomogeneous
Hagen-Poiseuille profiles and well describe the changes in recombination times
observed at T = 4 K and T = 25 K, respectively.%

\begin{figure}
\includegraphics[width=10cm]{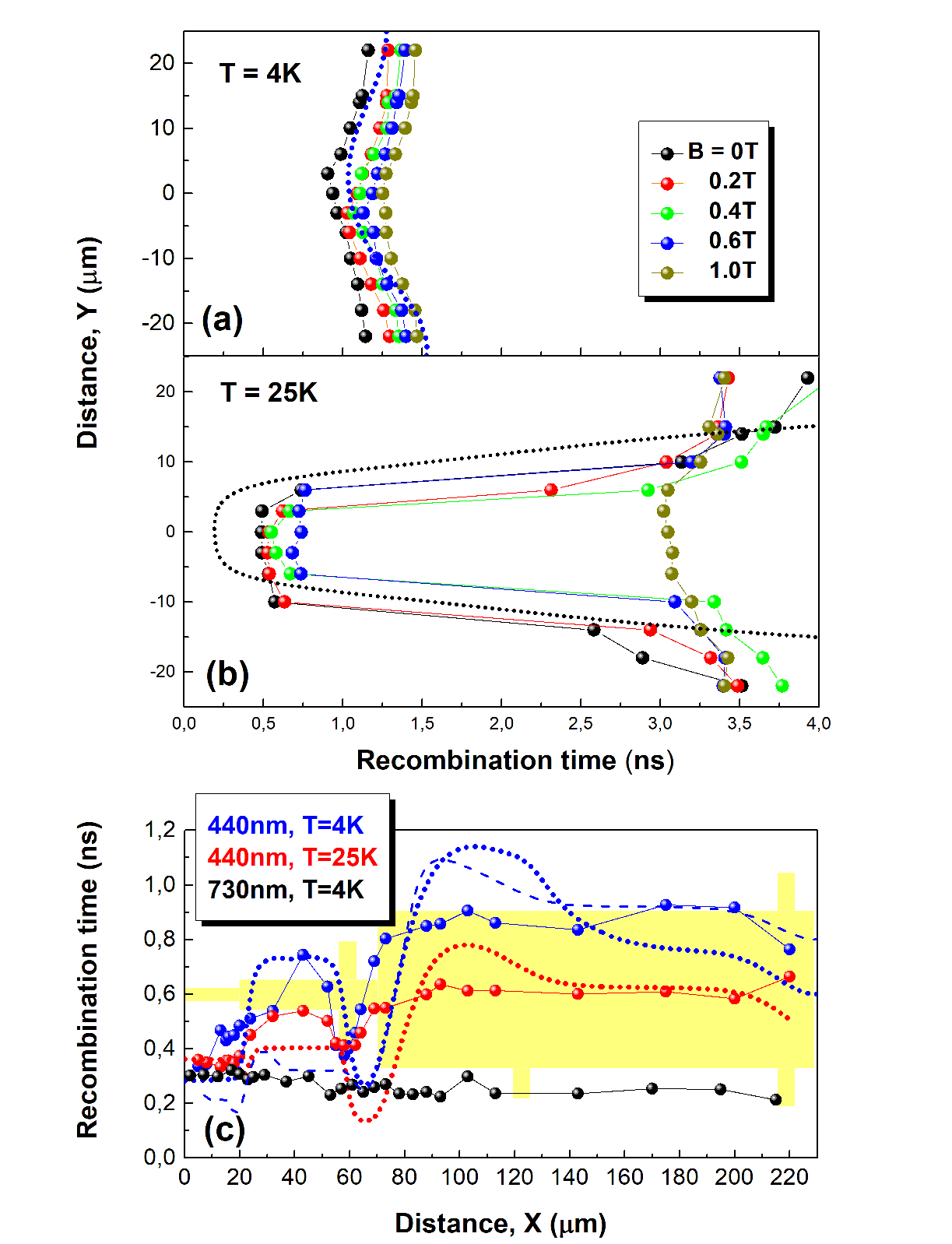}
\caption{(Color online) The recombination time mea-
sured perpendicular (a,b) and parallel (c) to the channel
at 4 K and 25 K, at an excitation energy of 2.33 eV
(440nm) and in various magnetic fields. Black dotted
lines were calculated as described in the text. The re-
combination time measured parallel to the channel at an
excitation energy of 1.7 eV (730nm) is shown in panel
(c). The measurements were carried out along the strips
shown in Fig. 2(b). Dotted are the recombination times
calculated for inviscid (blue line) and viscous (red line)
áuids, and as described in the text. Dashed line in panel
(c) calculated for 4 K including all contact pins. }
\end{figure}

Fig. 3 (color on-line). The recombination time measured perpendicular (a,b)
and parallel (c) to the channel at\ 4 K and 25 K, at an excitation energy of
2.33 eV (440nm) and in various magnetic fields. Black dotted lines were
calculated as described in the text. The recombination time measured parallel
to the channel at an excitation energy of 1.7 eV (730nm) is shown in panel
(c). The measurements were carried out along the strips shown in Fig. 2(b).
Dotted are the recombination times calculated for inviscid (blue line) and
viscous (red line) fluids, and as described in the text. Dashed line in panel
(c) calculated for 4 K including all contact pins.\bigskip\

Furthermore, a magnetic field creates a Lorentz force acting on the
hydrodynamic electron-hole fluid, which reduces the effective mean free path
and thereby suppresses the viscosity \cite{steinberg1958,alekseev2016}. The
data shown in Fig. 3(a,b) indicate that at T = 4 K the electron-hole fluid is
practically inviscid. In this case, the formation of Landau levels results in
a slight increase in the recombination time. At the same time, at T = 25 K,
when the electron-hole fluid is expected to be viscous, a strong change in the
time of recombination indicates a suppression of the viscosity induced by the
magnetic field. According to the data presented in Fig. 3(b), a 1 T magnetic
field transforms a viscous electron-hole liquid into an inviscid one with a
velocity profile similar to that obtained at T = 4 K.

The recombination time measured along the channel is depicted in Fig. 3(c). As
shown, under the high energy excitation of 2.33 eV, the recombination time
increases with increasing channel width. This is a direct consequence of the
Venturi effect: the recombination time is inversely proportional to the flow
rate of photogenerated holes, which decreases with increasing channel width.
To prove that the observed effect is due to the hydrodynamic flow of holes
injected from barriers, we measured the recombination times of holes
photogenerated by laser excitation with an energy of 1.7 eV, which creates
electron-hole pairs exclusively in the channel. In this case no significant
change in the recombination time is found. Thus, unlike holes injected from
barriers, electron-hole pairs photogenerated in a channel recombine rapidly
and do not form a hydrodynamic flow.

According to the Venturi effect, the flow velocity is inversely proportional
to the width of the channel sections. Thus, it is expected that the
recombination times measured in different sections will be proportional to the
section width and should be scaled as 1:2.5:12.5, where 1 corresponds to the
section with the minimum width. At the same time, the experimental
recombination times are scaled as 1:1.8:2.3 at T=4K and 1:1.5:1.9 at T = 25K.
It has been found that the presence of potentiometric contacts significantly
affects the flow velocity distribution. Adding contact pins considerably
reduces the corresponding flow velocity and allows the calculated
recombination time scale to be adjusted to the experimental one. The best fits
were obtained with contact pins to the middle and wide sections, as shown by
the dotted lines in Fig. 3(c). At the same time, the dashed line calculated
for an inviscid fluid shows that the addition of contact pins to a narrow
section somewhat worsens the agreement between the calculated and experimental
recombination times. The observed discrepancy is probably due to the fact that
the diffusion length of holes is comparable to the width of the narrow
section. In this case diffusion of holes is remarkably affected by the channel
boundary, which changes flow rate distribution. The effect of channel
boundaries on the recombination time of photogenerated holes was studied in
\cite{pusep2023}, where an optical analog of the Gurzhi effect was observed.

It should be noted that when calculating the flow rates and the corresponding
recombination times, no fitting parameters were used, except for the constant
C, which determines the relationship between the flow rate and the
recombination time and was taken equal to 6000 m and 4500 m for T = 4K
(inviscid flow) and 25K (viscous flow) respectively. Nevertheless, all the
observed experimental features of the recombination time are well reproduced.
This demonstrates the reliability of the model used.

The decrease in recombination time with temperature, shown in Fig. 3(c) is
associated with a change in the nature of the electron-hole fluid from
inviscid to viscous. The effect of temperature is more pronounced in the wide
section, and insignificant in the narrow one. The change in the recombination
time is associated with a change in the mean free path and, consequently, with
a change in the diffusion length $L$. The diffusion length of heavy holes as a
function of temperature was measured in a similar sample in \cite{pusep2022}.
Using these data, we compare the effect of temperature on the viscosity of an
electron-hole fluid observed by different independent methods. By using the
Einstein relation $D$ = $\mu E_{F}$/e, where $\mu$ is the mobility, and the
diffusion coefficient $D$ = $L^{2}/\tau$, the recombination time can be
expressed as:%

\begin{equation}
\tau=\frac{\mu E_{F}}{eL^{2}} \label{1}%
\end{equation}

If we assume that in the considered temperature range the mobility remains
unchanged, then the change in the recombination time is related to the
diffusion length. The diffusion lengths of heavy holes measured in
\cite{pusep2022} are 4 $\mu$m and 7 $\mu$m at 4K and 25K, respectively.
Application of the Eq. (1) to an electron-heavy hole fluid gives the
recombination time at 4 K 3 times longer than at 25 K, which somewhat exceeds
the change by 1.4 times observed in the wide section. However, in a viscous
fluid, mobility that increases with increasing temperature can improve the
agreement between calculated and measured recombination times. In a narrow
section, diffusion is limited by its width and, therefore, does not depend
noticeably on temperature.

In addition, the effect of the magnetic field on the distribution of
velocities along the channel, shown in Fig. 4 again reflects the influence of
the magnetic field on the viscosity of the electron-hole fluid. As shown in
Fig. 3(c), the difference in recombination times measured in channel regions
of different widths increases with decreasing viscosity. A similar increase in
the recombination time was found with increasing magnetic field. In the 4 $%
\mu
$m wide section of the channel, due to the higher diffusion velocity, the
effect of the magnetic field on viscosity is achieved at a stronger magnetic
field than in wide sections. In the 4 $%
\mu
$m wide section of the channel, due to the higher diffusion velocity, the
effect of the magnetic field on viscosity is achieved at a stronger magnetic
field than in wide sections.%

\begin{figure}
\includegraphics[width=10cm]{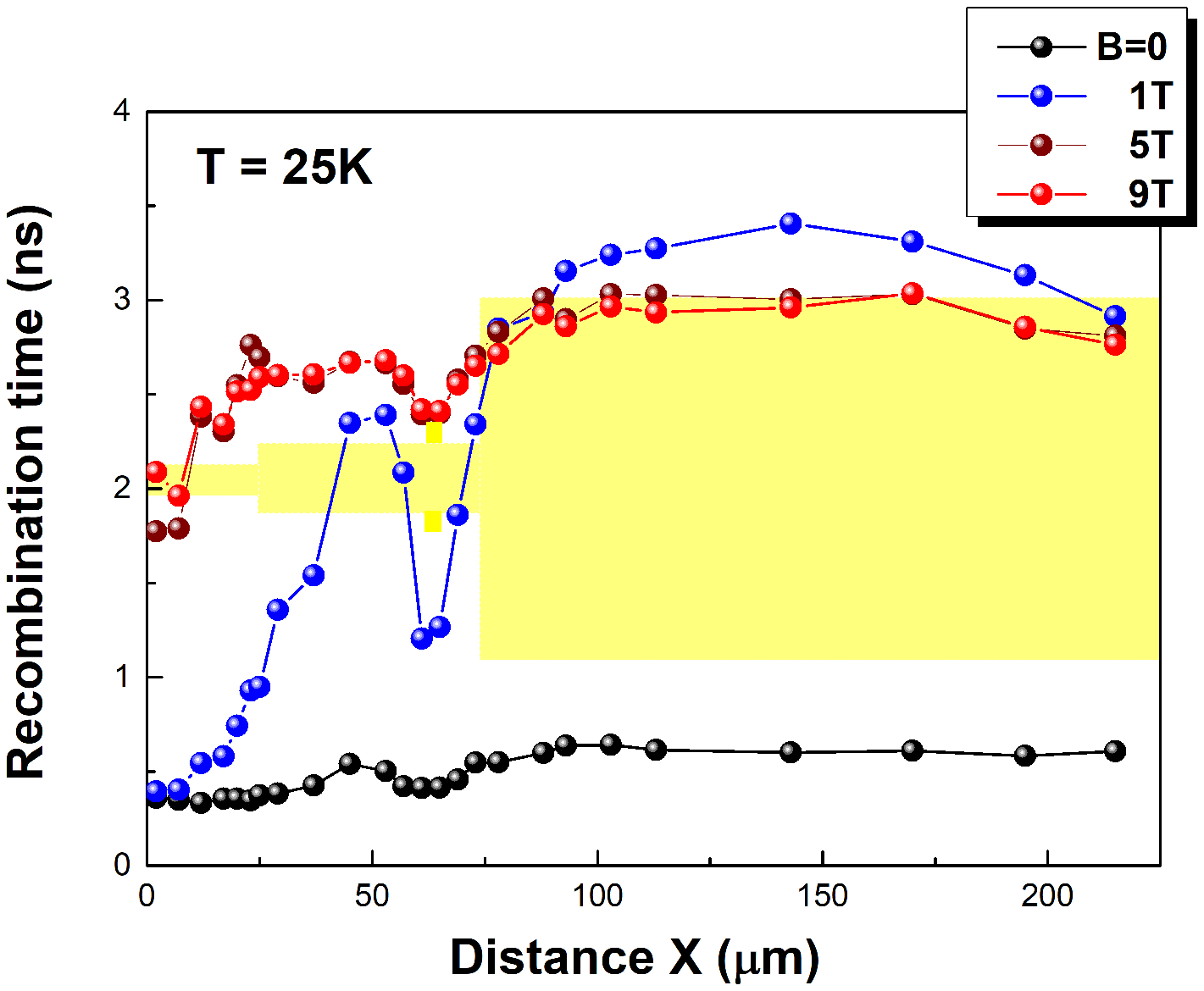}
\caption{(Color online) The recombination times measured along the channel at 25 K in various magnetic
fields. }
\end{figure}

Fig. 4 (color on-line). The recombination times measured along the channel
at\ 25 K in various magnetic fields.\bigskip

Taking into account the phenomenological character of the model used to
calculate the flow velocity distribution in the channel, a good agreement was
found between the experimental and calculated recombination times. This proves
the hydrodynamic nature of the diffusion of photogenerated holes in the sample
studied here. Despite its simplicity, the model used reproduces the main
experimental features, such as: change in the recombination time in sections
of different widths and across the channel, change in the recombination time
with a change in viscosity. Model calculations reproduce even the minimum of
the recombination time observed at the point where the channel width changes.

\section{Conclusion}

In summary, to study the diffusion of holes in a hydrodynamic electron system,
a new method for determining the diffusion velocity was used, which consists
in measuring the recombination rate of photogenerated holes. Accordingly, TRPL
scanning with micrometer resolution was performed in a mesoscopic
variable-width GaAs channel, where the hydrodynamic features of the
electron-hole fluid, such as the Venturi effect and the parabolic
Hagen-Poiseuille velocity profile, are demonstrated. The observed decrease in
the recombination time with increasing temperature is unusual and indicates
the viscous nature of the investigated electron-hole fluid. This effect is
similar to the decrease in the electrical resistance of viscous electrons with
increasing temperature. Moreover, a magnetic field induced suppression of
viscosity is observed. The presented results manifest to the fundamental role
of viscosity in the hydrodynamic flow of an electron-hole fluid in the GaAs
mesoscopic channel under study.

Acknowledgments: Financial supports from the Brazilian agencies FAPESP (Grant
2022/02132-0) and CAPES (Grant PNPD 88887.336083/2019-00) are gratefully acknowledged.


\begin{thebibliography}{99}                                                                                               %


\bibitem {geim2020}M. Polini, and A. Geim, Viscous electron fluids, Physics
Today \textbf{73}, 28 (2020).

\bibitem {fritz2023}L. Fritz, and T. Scaffidi, Hydrodynamic electronic
transport, arXiv:2303.14205.

\bibitem {white}F. M. White, Fluid Dynamics, McGraw Hill, 2009.

\bibitem {govorov2004}A. O. Govorov and J, J. Heremans, Hydrodynamic Effects
in Interacting Fermi Electron Jets, Phys. Rev. Lett. \textbf{92,} 026803 (2004).

\bibitem {taubert2010}D. Taubert, G. J. Schinner, H. P. Tranitz, W.
Wegscheider, C. Tomaras, S. Kehrein, and S. Ludwig, Electron-avalanche
amplifier based on the electronic Venturi effect, Phys. Rev. \textbf{B82},
161416 (R) (2010).

\bibitem {taubert2011}D. Taubert, G. J. Schinner, C. Tomaras, H. P. Tranitz,
W. Wegscheider, and S. Ludwig, An electron jet pump: The Venturi effect of a
Fermi liquid, J. Appl. Phys. \textbf{109}, 102412 (2011).

\bibitem {huang2023}W. Huang, T. Paul, K. Watanabe, T. Taniguchi, M. L.
Perrin, and M. Calame, Electronic Poiseuille flow in hexagonal boron nitride
encapsulated graphene field effect transistors, Phys. Rev. Research
\textbf{5}, 023075 (2023).

\bibitem {sulpizio2019}J. A. Sulpizio, L. Ella, A. Rozen, J. Birkbeck, D. J.
Perello, D. Dutta, M. Ben-Shalom, T. Taniguchi, K. Watanabe, T. Holder, R.
Queiroz, A. Principi, A. Stern, T. Scaffidi, A. K. Geim, and S. Ilani,
Visualizing Poiseuille flow of hydrodynamic electrons, Nature, \textbf{576},
75 (2019).

\bibitem {ku2020}M. J. H. Ku, T. X. Zhou, Q. Li, Y. J. Shin, J. K. Shi, C.
Burch, L. E. Anderson, A. T. Pierce, Y. Xie, A. Hamo, U. Vool, H. Zhang, F.
Casola, T. Taniguchi, K. Watanabe, M. M. Fogler, P. Kim, A. Yacoby, and R. L.
Walsworth, Imaging viscous flow of the Dirac fluid in graphene, Nature,
\textbf{583}, 537 (2020).

\bibitem {krebs2023}Z. J. Krebs, W. A. Behn, S. Li, K. J. Smith, K. Watanabe,
T. Taniguchi, A. Levchenko, V. W. Brar, Imaging the breaking of electrostatic
dams in graphene for ballistic and viscous fluids, Science \textbf{379}, 671 (2023).

\bibitem {Aharon2022}A. Aharon-Steinberg, T. V\"{o}lkl, A. Kaplan, A. K.
Pariari, I. Roy, T. Holder, Y. Wolf, A. Y. Meltzer, Y. Myasoedov, M. E. Huber,
B. Yan, G. Falkovich, L. S. Levitov, M. H\"{u}cker, E. Zeldov, Direct
observation of vortices in an electron fluid, Nature, \textbf{607}, 74 (2022).

\bibitem {braem2018}B. A. Braem, F. M. D. Pellegrino, A. Principi, M.
R\"{o}\"{o}sli, C. Gold, S. Hennel, J. V. Koski, M. Berl, W. Dietsche, W.
Wegscheider, M. Polini, T. Ihn, and K. Ensslin, Scanning gate microscopy in a
viscous electron fluid, Phys. Rev. \textbf{98} 241304(R) (2018).

\bibitem {pusep2022}Y. A. Pusep, M. D. Teodoro, V. Laurindo Jr., E. R. C. de
Oliveira, G. M. Gusev, and A. K. Bakarov, Diffusion of Photoexcited Holes in a
Viscous Electron Fluid, Phys. Rev. Lett. \textbf{128,} 136801 (2022).

\bibitem {pusep2023}Yu. A. Pusep, M. D. Teodoro, M. A. T. Patricio, G. M.
Jacobsen, G. M. Gusev, A. D. Levin, and A. K. Bakarov, Dynamics of
recombination in viscous electron--hole plasma in a mesoscopic GaAs channel,
J. Phys. D: Appl. Phys. \textbf{5}6, 175301 (2023).

\bibitem {dassarma2022}S. Ahn, and S. Das Sarma, Hydrodynamics, viscous
electron fluid, and Wiedeman-Franz law in two-dimensional semiconductors,
Phys. Rev. \textbf{B106}, L081303 (2022).

\bibitem {gusev2021}G. M. Gusev, A. S. Jaroshevich, A. D. Levin, Z. D. Kvon,
and A. K. Bakarov, Viscous magnetotransport and Gurzhi effect in bilayer
electron system, Phys. Rev. \textbf{B103}, 075303 (2021).

\bibitem {gusev2023}A. D. Levin, G. M. Gusev, A. S. Yaroshevich, Z. D. Kvon,
and A. K. Bakarov, Geometric engineering of viscous magnetotransport in a
two-dimensional electron system, Phys. Rev. \textbf{B108}, 115310 (2023).

\bibitem {steinberg1958}M. S. Steinberg, Viscosity of the Electron Gas in
Metals, Phys. Rev. \textbf{109}, 1486 (1958).

\bibitem {alekseev2016}P. S. Alekseev, Negative Magnetoresistance in Viscous
Flow of Two-Dimensional Electrons, Phys. Rev. Lett. \textbf{117,} 166601 (2016).
\end{thebibliography}
\end{document}